\documentclass{emulateapj}

\def\gsim{\;\rlap{\lower 2.5pt
 \hbox{$\sim$}}\raise 1.5pt\hbox{$>$}\;}
\def\lsim{\;\rlap{\lower 2.5pt
   \hbox{$\sim$}}\raise 1.5pt\hbox{$<$}\;}
\newcommand{\be}{\begin{equation}}
\newcommand{\beq}{\begin{equation}}
\newcommand{\ba}{\begin{eqnarray}}
\newcommand{\ee}{\end{equation}}
\newcommand{\eeq}{\end{equation}}
\newcommand{\ea}{\end{eqnarray}}

\usepackage{natbib}

\newcommand{\ga} {\gtrsim}
\newcommand{\la} {\lesssim}
\newcommand{\cel}{C_{E\ell}}
\newcommand{\ctl}{C_{T\ell}}
\newcommand{\ccl}{C_{C\ell}}
\newcommand{\Mpc}{{\rm Mpc}}

\begin{document}
\submitted{to appear in ApJ}
\title{The Reionization History at High Redshifts II:
Estimating the Optical Depth to Thomson Scattering from CMB
Polarization}

\author{Gilbert P. Holder}
\affil{School of Natural Sciences, Institute for Advanced Study, Princeton NJ 08540}
\author{Zolt\'an Haiman}
\affil{Department of Astronomy, Columbia University, 550 West 120th Street, New York
NY 10027}
\author{Manoj Kaplinghat, Lloyd Knox}
\affil{Department of Physics, 1 Shields Avenue, University of California, Davis,
CA 95616}

\begin{abstract}
In light of the recent inference of a 
high optical depth to Thomson scattering, $\tau$, from the WMAP data
we investigate the effects of extended periods of partial ionization 
and ask if the value of $\tau$ inferred by assuming a 
single sharp transition is an unbiased estimate.  
We construct and consider several representative ionization models and 
evaluate their signatures in the CMB.  If $\tau$ is estimated with a 
single sharp transition we show that there can be a significant bias 
in the derived value (and therefore a bias in $\sigma_8$ as well). 
For WMAP noise levels the bias in
$\tau$ is smaller than the statistical uncertainty, but
for Planck or a cosmic variance limited experiment the $\tau$
bias could be much larger than the statistical uncertainties.
This bias can be reduced in the 
ionization models we consider by fitting a slightly more complicated
ionization history, such as a two-step ionization process.  
Assuming this two-step process we find the 
Planck satellite can simultaneously determine the initial redshift
of reionization to $\pm 2$ and $\tau$ to $\pm 0.01$
Uncertainty about the ionization history appears to provide
a limit of $\sim 0.005$ on how well $\tau$ can be estimated
from CMB polarization data, much better than expected from WMAP but
significantly worse than expected from cosmic-variance limits.
\end{abstract}

\section{Introduction}

Understanding the reionization of the intergalactic medium is important for
cosmology for at least two reasons. How reionization occurred provides
crucial data on the first (and possibly second) generation of sources, 
while Thomson scattering of the photons of the cosmic microwave 
background (CMB) by free electrons suppresses ``primordial'' anisotropies 
in the CMB imprinted at $z \sim 1000$.

The first sources are not well understood \nocite{barkana01}
(see Barkana \& Loeb 2001 for a recent review), but there are several relevant
observational constraints. The observation of quasars at $z\sim 6$ that
show strong HI absorption \citep{becker01} indicates that the universe had
at least 1\% of the total hydrogen content in neutral form \citep{fan02}
at $z \sim 6$, with this neutral mass fraction rapidly decreasing at lower
redshifts (Songaila \& Cowie 2002).
This is a strong indication that the epoch of reionization ended at $z\sim6$.
On the other hand, the observed anisotropies of the CMB indicate that the
total optical depth to Thomson scattering is not extremely high, suggesting that
reionization couldn't have started at redshifts much higher than about 30
\citep{spergel03}.

What happened between redshifts 6 and 30 is unknown. There has been
extensive modeling and numerical simulation, but without a good understanding
of the sources robust conclusions are difficult to draw. Many studies have concluded
that reionization should happen fairly rapidly \citep{cen02,fan02}, 
but several recent studies have suggested 
(Wyithe \& Loeb 2003; Cen 2003; Haiman \& Holder 2003: hereafter paper I)
\nocite{wyithe03,cen03} 
that reionization is an extended process, perhaps even with 
multiple epochs of reionization. 

In recent work (Kaplinghat et al. 2003; hereafter K03) 
\nocite{kaplinghat03} it was shown that a two-step
reionization process could have an observable signature in the large-angle
CMB polarization anisotropies that would provide unique information on the
process of reionization in this difficult-to-probe redshift range. 
In this work, we extend these calculations to physically motivated
reionization histories provided by semi-analytic models and address the
possibility that non-trivial ionization histories can introduce a
bias in estimates of the optical depth $\tau$ if a simple one-step model
is used to fit the data.

The results from WMAP have opened a new window on the dark ages.
Several key cosmological parameters have been measured to high precision
and WMAP has observed the signature of free electrons at $z\sim 10$ for
the first time \citep{kogut03}. Previously, there had been practically no 
information about the ionization state of the IGM for $6.3 \la z \la 30$.
 
The optical depth to Thomson scattering is an important cosmological parameter. 
The temperature power spectrum only allows constraints on the
normalization of the primordial gravitational potential power spectrum
$P_\phi$ in the combination $P_\phi e^{-2\tau}$, so a determination of
$\tau$ allows a determination of the amplitude of potential (and mass) 
fluctuations.  There has been much recent controversy over the amplitude of 
mass fluctuations on scales of $8 h^{-1}$\,Mpc, $\sigma_8$, as summarized
in recent parameter estimates \citep{spergel03}. Estimates have varied by
nearly a factor of two between different methods of determination within
the last few years, so precise and accurate CMB estimates would be invaluable.
With accurate determinations of $\sigma_8$ it should be possible to put
strong constraints on the nature of the dark energy \citep{hu02}.

In the next section we describe the models that are used to generate
a zoo of ionization histories, while \S\ 3 demonstrates the effects of
non-trivial ionization histories on CMB polarization anisotropies and
parameter estimation. In \S\ 4 we provide a crude example of how it is
possible to reduce any biases induced by an unknown ionization history,
and we close with a discussion.

As a fiducial model we assume cosmological parameters $\Omega_m=0.29$,
$\Omega_{\Lambda}=0.71$, $\Omega_b\,h^2=0.024$, $h=0.72$ and an initial
matter power spectrum $P(k) \propto k$ in agreement with recent results
from WMAP \citep{spergel03}. 

\section{Reionization Models}
\label{sec:ion}

We employ semi--analytical models of reionization, in order to derive
a sample of physically motivated ionization histories.  Details of the
models are laid out in paper I, \nocite{haiman03} and we only
briefly review the main framework here.  The models assume that the
total volume fraction of ionized regions is being driven by ionizing
sources located in dark matter halos whose abundance is described by
the N--body simulations of Jenkins et al. (2001) We distinguish dark
matter halos in three different ranges of virial temperatures, as
follows:
\begin{center}
\begin{tabbing}
\hspace{2cm} \= $100\,{\rm K}$   $\lsim$ \= $T_{\rm vir}$ \=  $\lsim  10^4\, {\rm K}$ \hspace{1cm} \= (Type II)\\
             \> $10^4\,{\rm K}$  $\lsim$ \> $T_{\rm vir}$ \>  $\lsim 2\times 10^5\,{\rm K}$        \> (Type Ia)\\
             \>                          \> $T_{\rm vir}$ \>  $\gsim 2\times 10^5\,{\rm K}$        \> (Type Ib)
\end{tabbing}
\end{center}

We will hereafter refer to these three different types of halos as
Type II, Type Ia, and Type Ib halos.  Each type of halo plays a
different role in the reionization history.  In short, Type II halos
can host the first ionizing sources, but only in the neutral regions
of the IGM, and only if ${\rm H_2}$ molecules are present in
sufficient quantity to allow efficient cooling; Type Ia halos can only
form new ionizing sources in the neutral IGM regions, but irrespective
of the ${\rm H_2}$ abundance, and Type Ib halos can form ionizing
sources regardless of the ${\rm H_2}$ abundance, and whether they are
in the ionized or neutral phase of the IGM.

The contribution of each halo to reionization is quantified by
explicitly computing the expansion of the ionized Str\"omgren region,
dictated by the source luminosity and the background IGM density and
the clumping factor $C_{\rm HII}$.  We allow the three different
sources above to have different efficiencies $\epsilon$ of injecting
ionizing radiation into the IGM.  Here $\epsilon\equiv N_\gamma f_{\rm
esc}f_*$, where $f_* \equiv M_*/(\Omega_{\rm b}M_{\rm halo}/\Omega_m)$
is the fraction of baryons in the halo that turns into stars ($\sim
10\%$ in normal stars $\lsim 0.01$ in Type II halos); $N_\gamma$ is
the mean number of ionizing photons produced by an atom cycled through
stars, averaged over the initial mass function (IMF) of the stars
($\sim 4000$ for a normal Salpeter IMF, and up to a factor of 20 higher
for a population of massive, metal--free stars (Bromm, Kudritzki \&
Loeb 2001; Schaerer 2002); and $f_{\rm esc}$ is the fraction of these
ionizing photons that escapes into the IGM ($\sim 10\%$ for Types Ia,b
halos, and $\sim 1$ for Type II halos). In our models, we also allow
the possibility that radiative feedback effects photo--dissociate
${\rm H_2}$ molecules below some critical redshift $z_{\rm uv}$, and
we self--consistently exclude the Type II and Type Ia halos from
forming any ionizing sources inside regions that had already been
reionized.  We adopt a fixed $C_{\rm HII}=10$ in all models; variations
in the clumping factor can be absorbed into changes in the
efficiencies with redshift.

In summary, our model has four parameters: the overall efficiencies,
$\epsilon_{\rm II}$, $\epsilon_{\rm Ia}$, $\epsilon_{\rm Ib}$, and the
redshift $z_{\rm uv}$ at which ${\rm H_2}$ dissociative feedback sets
in. We here use 5 different models that are broadly representative.
Three of the models were chosen to have optical depth equal to that
measured by the WMAP experiment \citep{kogut03} and two others were
chosen to investigate the effects of larger or smaller optical depths.
Details of the models can be found in paper I.

Model 1 assumes that for massive halos the stellar IMF is not
metal-free (pop II) while minihalos (cooled by $H_2$) form metal--free
stars that produce $\sim 20$ times more ionizing photons.  In terms of
model parameters, this corresponds to $(\epsilon_{\rm
HII},\epsilon_{\rm Ia,Ib})=(200,80)$.  It is assumed that $H_2$ starts
to be destroyed at $z_{\rm uv}=17$.  Model 2 assumes that minihalos do
not contribute to reionization (efficient destruction of $H_2$) and
that the efficiency in larger halos is increased to $\epsilon_{\rm
Ia}\sim 4800$.  In Model 3 it is assumed (as in Wyithe \& Loeb 2003;
Cen 2003) that there is a sharp transition from metal-free to normal
stars at $z=14$. Model 4 assumes that minihalos are more effective at
forming stars than in Model 1, with $\epsilon_{\rm II}=1400$ and has
molecules being destroyed at $z_{\rm uv} \sim 25$.  Finally, Model 5
assumes that feedback from star formation becomes efficient at
destroying molecules at $z_{\rm uv} \lesssim 21$ but the same
efficiencies as Model 1.

The ionization histories in our models are shown in Figure~1. As
discussed in depth in paper I, the physics of reionization is rich in
features that can naturally lead to distinctive ionization histories.
These features can arise because of (1) the different types of
coolants in halos with virial temperatures above and below
$\sim10^4$K, (2) the different response of different halos to
radiative feedback on the ${\rm H_2}$ chemistry, and to
photoionization feedback on gas infall, and (3) the different
properties of metal--free and normal stellar populations.

In all models, we assume singly-ionized Helium traces ionized hydrogen
(i.e., 1.08 free electrons per hydrogen atom for a completely ionized
universe).  In what follows, discussion of $x_e$ ignores Helium (e.g.,
complete ionization is referred to as $x_e = 1$), but the factor of
1.08 is included in all calculations.

\section{Large Angle CMB Polarization Anisotropies}
\label{sec:cmb}

We modified CMBFast\footnote{available at 
http://www.cmbfast.org} 
\citep{seljak96} to use the ionization histories from the previous section
to generate temperature (TT), polarization (EE), and 
cross anisotropy (TE) power spectra,
$\ctl,\cel$ and $\ccl$ respectively. 
A similar modification was done by \citet{bruscoli02} but for an ionization
history extracted from a numerical simulation, and by 
\citet{naselsky03} using different ionization histories. 
The ionization histories are shown in Figure 1 and the corresponding 
power spectra are shown in Figures 2 and 3. 

Note in Figure 2 that for larger optical depths, there are secondary bumps in
the $C_{El}$.  These are thoroughly explained in \citet{zaldarriaga97}.
For the $C_{Cl}$ (shown in Figure 3)
an important difference is that we are correlating
quantities which, for each value of $k$, have different angular
frequencies on the sky.  The polarization has angular frequency
$l = k(\eta_0 - \eta_{ri})$ (where $\eta_0$ and $\eta_{ri}$ are the
conformal times today and at the onset of reionization) 
since it is projecting from the epoch
of reionization where it was created.  The temperature has a
correspondingly
higher angular frequency since it is projecting from the (further)
last-scattering surface.  The matched angular frequencies of $E$
correlated with $E$ lead to secondary peaks in $C_{El}$, whereas the mismatched
angular
frequencies of $T$ and $E$ wash out the fluctuation power and do
not lead to secondary peaks in $C_{Cl}$.  

As found in K03, for highly sensitive experiments approaching
the cosmic-variance limit, almost all the sensitivity to $\tau$ comes from
$C_{El}$.  This is because $C_{El} \propto \tau^2$ whereas
$C_{Cl} \propto \tau$ and because the fractional uncertainty in
$C_{El}$ is smaller than the fractional uncertainty in $C_{Cl}$ in
the cosmic variance limit.  These fractional uncertainties would
be equal in the limit of perfect correlation ($C_{Cl} =
\sqrt{C_{El}C_{Tl}}$).

We normalize $P_\Phi(k=0.05\Mpc^{-1}) e^{-2\tau}$ by requiring that
the temperature fluctuation at $\ell=150$ is 150 $\mu K$.  This
choice is arbitrary, but largely irrelevant for our purposes. 
Varying the ionization history with this product fixed produces no change 
in the angular power spectra
at $\ell \ga 50$ (except due to non-linear effects at $\ell \ga 2000$).  
Note that since $\ctl$ has been measured well in this range,
a higher optical depth requires a larger normalization of $P_\Phi$ 
in order to agree with the data.  As outlined in K03, variations 
in the fiducial model, such as a slight tilt or a slightly
different normalization, will not have a 
large effect on our conclusions. 

\begin{figure}
\plotone{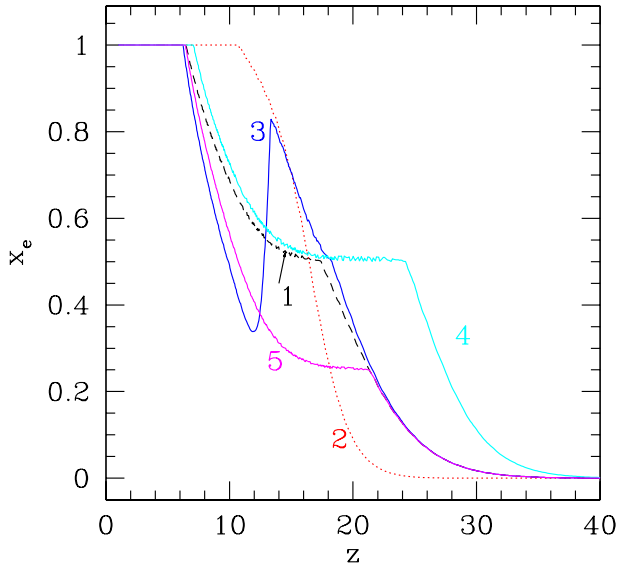}
\caption{Ionization histories for the five models (explained in the text) 
considered.} 
\label{fig:models} 
\end{figure}

\begin{figure}
\plotone{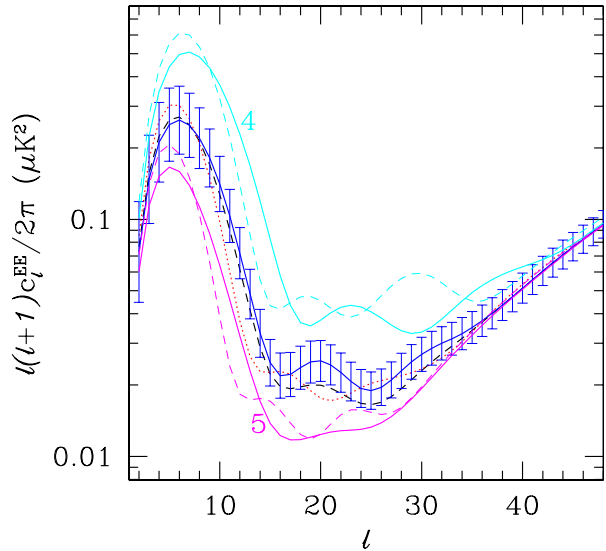}
\caption{Polarization 
power spectra for the five models (see Figure 1) considered. 
Model 1 is shown as a bold dashed line, model 2 is shown as a dashed line
and model 3 is the solid line with cosmic variance error bars shown. Note
that models 1--3 all have the same optical depth.
For models 4 and 5 the best fit single transition model polarization spectra 
are shown as dashed lines of the same color. 
All models are normalized to give the same temperature
power spectrum at $\ell \ga 50 $. 
} 
\label{fig:models2} 
\end{figure}

\begin{figure}
\plotone{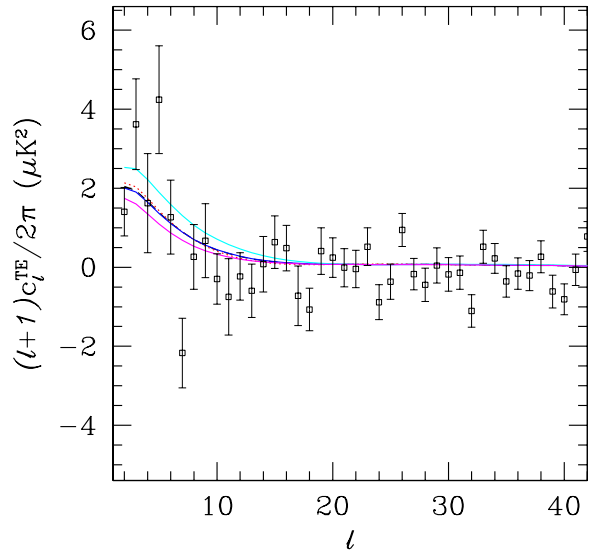}
\caption{Same five models as in Fig. 2 but now for the cross--correlation
and with WMAP measurements plotted. As before, top line is model 4, bottom
line is model 5, and models 1--3 are nearly indistinguishable.
} 
\label{fig:models2cross} 
\end{figure}

To explore questions of bias in $\tau$ we see how well purely 
phenomenological models with one or two sharp transitions can be used
to fit our physical models (1)-(5).  For measurement uncertainties
we assume Gaussian, white detector noise and ignore
beam effects.  We calculate the likelihood of the
phenomenological models, given one of the physical models as the ``data'',
denoted now with a $d$ superscript (see K03):
\begin{eqnarray}
\bar{\chi}^2 \equiv -2 \ln\mathcal{L} = \sum_{\ell} (2 \ell + 1) f_{sky} 
\Bigl [ \ln ([\cel+w^{-1}] \ctl -\ccl^2)  \nonumber \\
+ { [\cel+w^{-1}] \ctl^{d} + \ctl \cel^{d} - 2\ccl \ccl^{d} \over 
[\cel+w^{-1}] \ctl - \ccl^2} \Bigr ] \quad .
\end{eqnarray}
In the above, $f_{sky}$ is the fraction of sky coverage (which we take to
be unity), and $w$ is the weight per unit solid angle for polarization
measurements. We have assumed that for our purposes we can ignore detector
noise in the measurement of the temperature power spectrum. Note that we
have included the effects of all the data, i.e., EE, TT, and TE correlations
are all implicitly taken into account. We have not included possible 
effects of foregrounds. For WMAP, we assume $w=2\times 10^{14}$, roughly
the expected two-year two channel sensitivity (allowing other frequencies
to be used for foreground removal) while for Planck we assume 
$w=1.67\times 10^{16}$, corresponding roughly to one year and two
frequencies. 
We adopt these values as representative of the expected performance of the
instruments, but should be viewed as order of magnitude estimates.
For an ideal cosmic variance limited experiment $w^{-1}=0$.
We require that $\ctl > 0.01 w^{-1}$ in the fiducial model to be
included in the likelihood calculation to suppress contributions from 
points with low signal-to-noise.

\begin{table*}[hbt]\small
\caption{\label{table:sharp}
Best fit single-step reionizations 
(roughly, number of $\sigma$ is $\sqrt{\Delta \bar{\chi}^2}$)}
\begin{center}
\begin{tabular}{lr|rrr|rrr|rrr}
 &  &  &  Cos. Var.& & &  & WMAP & & & Planck\\
\hline
model & $\tau$ & $z_{tr}$ & $\tau_{cv}$ & $\Delta \bar{\chi}^2_{cv}$ &
$z_{tr}$ & $\tau_{wmap}$ & $\Delta \bar{\chi}^2_{wmap}$ &
$z_{tr}$ & $\tau_{planck}$ & $\Delta \bar{\chi}^2_{planck}$ \\
\hline
1 & 0.169 & 16.3 & 0.166 & 57 &16.1 & 0.163 & 0.3 & 16.9 &0.174 & 15 \\
2 & 0.169 & 16.1 & 0.163 & 9 & 16.3  & 0.166 & 0.0& 16.3 &0.166 & 2 \\
3 & 0.169 & 17.0 & 0.176 & 49 & 16.2 & 0.164 & 0.4& 17.3 &0.181 & 16 \\
4 & 0.228 & 20.4 & 0.229 & 112& 19.6 & 0.216 & 1.1& 20.9 &0.238 & 39 \\
5 & 0.139 & 14.4 & 0.138 & 43 & 13.8 &0.130  & 0.2& 14.9 &0.145 & 13 \\
\hline
\multicolumn{5}{l}{} \\
\end{tabular}\\[12pt]
\begin{minipage}{5.2in}
\end{minipage}
\end{center}
\end{table*}

For simplicity, we only include terms up to $\ell=50$. There is practically
no information in higher multipoles, although there is likely to be 
some signature at $\ell \ga 2000$ from non-linear effects, which could be
important.
We minimize this function by adjusting the transition redshift, $z_{tr}$, of 
a model with sudden reionization and calculate the difference in
$\bar{\chi}^2$ of this best-fit sudden model relative to the true
model. The true model is thus $\exp(\Delta \bar{\chi}^2/2)$
more likely than the most likely phenomenological model.
For Gaussian statistics, our estimator $\bar{\chi}^2$ is equal to
the usual $\chi^2$ statistic, so a rough estimate of the number of
``sigmas'' is $\sqrt{\Delta \bar{\chi}^2}$.  Large values of this misfit
statistic indicate that the true model is a much better fit than the
model being considered while small values indicate a model that
is virtually indistinguishable from the input model. The best fitting
sudden models are indicated in Table 1, along with the difference in
$\bar{\chi}^2$ and the optical depth of both the best fit and the 
input model.  
The first two columns indicate model number (see Figure 1) and true optical 
depth. Columns 3-5 show results of fitting a single sharp reionization 
assuming cosmic variance error bars and indicate the best fit single 
reionization redshift, 
best fit optical depth, and difference in $\bar{\chi}^2$ relative
to true model. Columns 5, 6 and 7 show best fit single transition
redshift, optical depth and misfit
statistic assuming WMAP noise levels, while the last three columns show
the same parameters assuming Planck noise levels.

For some of the models the misfit is very large, in one case a shift in the
misfit statistic of more than 100 (roughly ``10 $\sigma$'') for a cosmic
variance limited experiment. This confirms 
that there is significantly more information in the large angle
polarization signal than simply the optical depth, as shown in K03. For the
most basic reionization signal the misfit is just above the ``3 $\sigma$''
level for cosmic variance limits, 
indicating that if reionization happens fairly quickly the exact
nature of the transition is unimportant and the dominant effect on the
CMB signal will be only that of the optical depth. This corresponds to the
case studied by \citet{bruscoli02}, although even for this case our results
are slightly less pessimistic. Part of this is due to the higher optical
depth of our fiducial model, providing more signal.

From the point of view of parameter estimation, it is striking that the
best fit sharp transition can provide a biased estimate of the optical
depth, especially compared to the statistical uncertainties.  For the
assumed sensitivity of WMAP the statistical uncertainty $\delta \tau$ should be $\sim 0.02$,
for Planck $\delta \tau \sim 0.005$ and for cosmic variance errors
$\delta \tau \sim 0.002-0.003$.
For the more exotic ionization histories the optical depth can be
seen to be significantly biased for cosmic variance level measurements, 
with the direction and the magnitude of the 
bias sensitive to the details of the ionization history. 
Using the incorrect ionization history for model fitting introduces
a systematic error in the value of $\tau$ with a direction and magnitude 
that depends on the details of $x_e(z)$.
As seen in models 3 and 4, offsets between
the true and derived optical depths could easily be $\gtrsim$0.01 for Planck.
Determination of the optical depth, and
thus the matter power spectrum amplitude, will be limited by a lack
of understanding of the nature of the reionization process.
For WMAP, it appears
that the derived optical depth assuming a single sharp
transition will not be highly biased for any
of the reionization models that we consider, given that the estimated
uncertainty in $\tau$ is $\sim 0.03$ (K03). 

\section{Toward Unbiased Optical Depth Estimates}

There is information in the shape of the large angle polarization power
spectra, so it is informative to see what can be gleaned. In Figure
4 we show the results of an analysis similar to that of K03. 
We used a simple two-step model, where it is assumed that a transition
from partial ionization to full ionization occurred 
at $z=6.3$ and that a transition
from nearly zero to an intermediate (constant) ionization fraction, $x_e$, 
occurred at $z=z_e$. We examine two fiducial models,
one with $z_e=25$ and the other
with $z_e=16$, with $x_e$ values chosen so that both have 
$\tau = 0.148$.
We then investigate the accuracy with which $\tau$ and $z_e$ 
can be recovered assuming noise levels typical for WMAP or expected 
for Planck by once again taking the fiducial model as the data 
and evaluating the likelihood (Eq. 1) as a function of the two
parameters of the model.  Figure 4 displays contours of constant
likelihood for the three experimental cases.
At noise levels appropriate for WMAP
it will be very difficult to differentiate between different models that
yield the same optical depth (as pointed out in K03 and 
verified by Kogut et al. 2003), but Planck would be able to determine the
onset of partial reionization quite well and a cosmic variance limited
experiment would be able to determine this onset very precisely.
There will therefore likely be suggestions in the data
itself pointing to better models for measuring $\tau$.

\begin{figure}[tbh]
\plotone{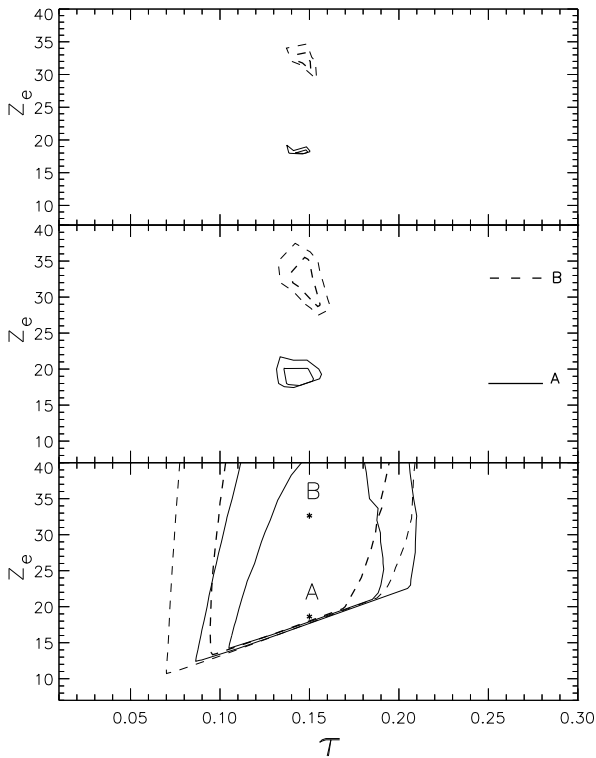}
\caption{Constant likelihood contours at 10\% (thick curves)
and 1\% (thin curves) of the maximum likelihood -- which
occurs at the fiducial models labeled by A and  B denoted by
asterisks. A is a model completely reionized below $z=18$
and B has $z_{\rm early}=32$ and $x_e=0.4$; both models have 
$\tau=0.148$. All the panels have the same fiducial models.
The solid curves correspond to model A, the dashed curves to
model B. The upper panel is for a cosmic variance limited
experiment (about 25 times the raw sensitivity of Planck).
The middle panel assumes 1 Planck channel sensitivity while
the lower panel is for two channel WMAP sensitivity. }
\label{fig:cont-tau}
\end{figure}

In the previous section, we saw that the largest biases arise in the cases 
where the misfit statistic is also large. Therefore when the 
measured $\tau$ is highly biased the quality of fit will probably also be 
bad, indicating a possibly contaminated result. We now investigate whether
a slightly more complicated fitting form for the ionization history can
lead to better estimates of the optical depth.
As a simple example of a path to a possibly less 
biased estimate of the optical depth, we fit models with a two-step
reionization process, where the ionization history is characterized by a
redshift of first ionization $z_{early}$, when the ionized fraction went
quickly from effectively zero to $x_e$, and a second redshift, 
$z_{late}$ when the ionized fraction went quickly to unity. 
To ensure stability in the numerical implementation we have jumps
in the ionization fraction take place over a range in redshift of 
$\Delta z=1$ centered on the nominal redshift of the transition and interpolated
in $log(z)$.
For each ionization model from the previous section, we vary
the three ionization parameters to find the values that minimize the 
misfit statistic, assuming cosmic variance errors only.

\begin{table}[tbph]
\label{table:twostep}
\begin{center}
\caption{Best fit two-step reionizations, assuming cosmic variance 
limited errors} 
\vspace{.1in}
\begin{tabular}{lr|rrrrr|r}
model & $\tau$ & $z_{late}$ & $z_{early}$ & $x_e$ & $\tau_{cv}$ &
$\Delta \bar{\chi}^2_{cv}$ & $\delta \tau$ \\
\hline
1 & 0.169 & 10 & 24 & 0.45 & 0.172 & 3   & $\pm$ 0.006  \\ 
2 & 0.169 & 14 & 20 & 0.45 & 0.169 & 0.2 & $\pm$ 0.004  \\ 
3 & 0.169 & 8  & 23 & 0.55 &  0.171 & 2  & $\pm$ 0.005  \\ 
4 & 0.228 & 18 & 29 & 0.30 & 0.234 & 3   & $\pm$ 0.008  \\
5 & 0.139 & 3  & 34 & 0.65 & 0.140 & 2   & $\pm$ 0.004  \\
\hline
\multicolumn{5}{l}{} \\
\end{tabular}
\end{center}
\end{table}

Table 2 shows the derived optical depths from fitting a two-step model.
Column 2 indicates the true optical depth, the third column
indicates redshift at which $x_e=1$ and the fourth column shows redshift
at which $x_e$ changes from effectively zero at higher redshift to
the value shown in column 5. The fifth column shows optical depth of this
best fit model, while the sixth column shows the difference in 
$-2ln\mathcal{L}$ of this best fit relative to the true model.
The final column shows the uncertainty in the determination
of $\tau$ when the extra parameters are allowed in the fit. This
uncertainty was determined from the smallest and largest values
of $\tau$ found in models with $\Delta \bar{\chi}^2 <1$.

As can be seen in Table 2, in this case the misfit
statistics are much lower and the optical depth estimates are much less
biased.  In most cases the estimated values of $\tau$ are now biased at levels
near or below the cosmic variance statistical errors, indicating that
imperfect knowledge of the ionization history is not a fundamental limit
to a good measure of $\tau$. In model 4 our
two-step model may not be adequate, in that the bias in $\tau$ is comparable
to the statistical uncertainty. If the
measured optical depth is very high ($\ga 0.2$) some care will be required
to obtain a precise and accurate estimate of the amplitude of the matter power
spectrum. Note that the reduction in bias
has come with the cost of statistical errors in $\tau$ increasing 
by a factor of $~2$.

In general, ionization histories with widely separated (in redshift)
episodes of ionization seem most prone to biased estimates of $\tau$.
We have not provided an exhaustive exploration of the possible
parameter space of ionization histories, so it is still
possible that reionization histories exist with larger biases 
and/or there are cases where a two-step
reionization history does little to improve estimates of $\tau$.

It is likely that there is a more physical parameterization 
of the ionization history that can both minimize
biases in parameter estimates and provide insight into the first generation
of sources. If the optical depth is measured to be higher than 0.1, as
hinted by recent WMAP results, it will be important to find a good 
parameterization.

\section{Discussion}
\label{sec:disc}

We have shown that large angle polarization measurements could be
very useful for shedding light on the end of the dark ages, a topic
addressed in further detail in paper I. Conversely,
it appears that at least a rudimentary understanding of the dark ages,
beyond a simple optical depth to some characteristic redshift,
will be required to be able to measure the amplitude of primordial
fluctuations to very high accuracy. Particularly if CMB measurements
suggest that the optical depth is high ($\ga 0.2$), there is a real
danger of mis-estimating the true optical depth by a significant 
amount ($>0.01$). In the near term, it appears that estimates of
the optical depth based on the WMAP satellite will not be heavily biased
at the level of the expected precision of 0.03 (K03). Therefore, the
derived values of $\sigma_8$ in recent work are robust (
within the statistical uncertainties) to the choice of
ionization history that is used to do the fit. This is unlikely to be
the case for the next generation of instruments, given the apparently
high optical depth measured by WMAP.

Allowing even a moderately more complex ionization history allows much
of the bias to be removed, at the expense of larger uncertainties. 
Uncertainty in the ionization history appears to provide a 
floor of $\sim 0.005$ on how well we can measure the optical
depth to Thomson scattering from CMB polarization observations.

Foreground contamination of large angle measurements is currently
unknown for polarization experiments. With its extensive frequency
coverage, Planck will be an exquisite instrument for characterizing
and assessing the importance of astronomical sources of polarization.
We believe we have been fairly conservative in our estimates of how
many frequencies will be available for CMB measurement.

CMB polarization measurements provide a unique complement to absorption
studies. While hydrogen absorption studies are sensitive to the fraction
of neutral hydrogen, CMB polarization is sensitive to the fraction of
ionized hydrogen. The signature of partial ionization at $z \sim 15$
will be very difficult
to detect using absorption studies, so it will be extremely useful to have
a complementary tool. 

\acknowledgements{
We are grateful to Wayne Hu, Avi Loeb, and Matias Zaldarriaga for useful 
discussions. GPH is supported by the W.M. Keck foundation. LK and MK were
supported by NASA grant NAG5-11098. We thank Uros Seljak and Matias
Zaldarriaga for the use of their CMBFast code.
}


\begin{thebibliography}{}

\bibitem[{Becker} {\em et~al.} (2001)]{becker01}
{Becker}, R.~H. {\em et~al.} 2001, \aj, {122}, 2850--2857.

\bibitem[Bromm, Ferrara, Coppi, \& Larson(2001)]{bromm01} 
Bromm, V., Ferrara, A., Coppi, P.~S., \& Larson, R.~B.\ 2001, \mnras, 328, 
969 

\bibitem[Bromm, Kudritzki, \& Loeb(2001)]{bromm01a} Bromm, V., 
Kudritzki, R.~P., \& Loeb, A.\ 2001, \apj, 552, 464 


\bibitem[{Bruscoli}, {Ferrara}, and {Scannapieco} (2002)]{bruscoli02}
{Bruscoli}, M., {Ferrara}, A., and {Scannapieco}, E. 2002, \mnras, {330},
  L43--L47.

\bibitem[{Cen} (2003)]{cen03}
{Cen}, R. 2003, \apj, { submitted}, astro--ph/0210473.

\bibitem[{Cen} and {McDonald} (2002)]{cen02}
{Cen}, R. and {McDonald}, P. 2002, \apj, { 570}, 457--462.


\bibitem[{Fan} {\em et~al.} (2002)]{fan02}
{Fan}, X., {Narayanan}, V.~K., {Strauss}, M.~A., {White}, R.~L., {Becker},
  R.~H., {Pentericci}, L., and {Rix}, H. 2002, \aj, { 123}, 1247--1257.


\bibitem[{Haiman} and {Holder} (2003)]{haiman03}
{Haiman}, Z., \& Holder, G.~P. 2003, ApJ, submitted




\bibitem[{Hu} (2002)]{hu02} Hu, W.\ 2002, \prd, 66, 83515

\bibitem[{Jenkins} {\em et~al.} (2001)]{jenkins01}
{Jenkins}, A., {Frenk}, C.~S., {White}, S.~D.~M., {Colberg}, J.~M., {Cole}, S.,
  {Evrard}, A.~E., {Couchman}, H.~M.~P., and {Yoshida}, N. 2001, \mnras, { 321}, 372.

\bibitem[{Kaplinghat} {\em et~al.} (2003)]{kaplinghat03}
{Kaplinghat}, M., {Chu}, M., {Haiman}, Z., {Holder}, G.~P., {Knox}, L., and
  {Skordis}, C. 2003, \apj, {583}, 24--32.

\bibitem[Kogut {\em et~al.} (2003)]{kogut03} 
Kogut, A. et al. 2003, ApJ, 
submitted, astro-ph/0302213

\bibitem[{Loeb} and {Barkana} (2001)]{barkana01}
{Loeb}, A. and {Barkana}, R. 2001, \araa, { 39}, 19--66.

\bibitem[{Naselsky} and {Chiang} (2003)]{naselsky03}
Naselsky, P. and Chiang, L.-Y. 2003, \mnras, {submitted}, astro-ph/0302085

\bibitem[Schaerer(2002)]{schaerer02} Schaerer, D.\ 2002, \aap, 
382, 28 

\bibitem[{Seljak} and {Zaldarriaga} (1996)]{seljak96}
{Seljak}, U. and {Zaldarriaga}, M. 1996, \apj, { 469}, 437

\bibitem[{Songaila} \& {Cowie} (2002)]{songaila02}
Songaila, A., \& Cowie, L.L. 2002, \aj, 123, 2183

\bibitem[{Spergel} et al. (2003)]{spergel03} Spergel, D. N. et al. 2003,
ApJ, submitted, astro-ph/0302209

\bibitem[{Wyithe} and {Loeb} (2003)]{wyithe03}
{Wyithe}, S. and {Loeb}, A. 2003, \apj, {submitted}, astro--ph/0209056.

\bibitem[Zaldarriaga(1997)]{zaldarriaga97} Zaldarriaga, M.\ 1997, 
\prd, 55, 1822 


\end{thebibliography}
\end{document}